\documentstyle[preprint,aps]{revtex}
\newcommand{\be}{\begin{equation}}
\newcommand{\ee}{\end{equation}}
\newcommand{\ber}{\begin{eqnarray}}
\newcommand{\eer}{\end{eqnarray}}

\newcommand{\la}{{\langle}}
\newcommand{\ra}{{\rangle}}

\input psfig.sty
\begin{document}
\draft

\title{Quasiparticle Lifetime in a Finite System: A Non--Perturbative
Approach}

\author{
Boris~L.~Altshuler$^1$, Yuval~Gefen$^2$,
Alex~Kamenev$^2$, and Leonid~S.~Levitov$^3$}

\address{
$^1$NEC Research Institute, 4 Independence way, Princeton, NJ 08540,\\
$^2$Department of Condensed Matter Physics,
The Weizmann Institute of Science, Rehovot, 76100, Israel,\\
$^3$Massachusetts Institute of Technology, 12-112, Cambridge, MA 02139 }
\maketitle

\begin{abstract}
  The problem of electron--electron lifetime in a quantum dot is
studied beyond perturbation theory by mapping it onto the problem
of localization in the Fock space. We
identify two regimes, localized and delocalized,
corresponding to quasiparticle spectral peaks of zero and finite
width, respectively. In the localized regime,
quasiparticle states are very close to single particle excitations.
In the delocalized state, each eigenstate is a
superposition of states with very different quasiparticle
content. A transition between the two regimes occurs at the
energy $\simeq\Delta(g/\ln g)^{1/2}$, where $\Delta$ is the one
particle level spacing, and $g$ is the dimensionless
conductance. Near this energy there is a broad critical region
in which the states are multifractal, and are not described by
the Golden Rule.
   \end{abstract}


Quasiparticle in a Fermi-liquid is not an eigenstate: it decays
into two quasiparticles and a hole. In an infinite clean system,
by using the Golden Rule (GR), quasiparticle decay rate is estimated
as $\gamma(\epsilon)\sim {\epsilon^2}/\epsilon_F$, where
$\epsilon$ is quasiparticle energy and $\epsilon_F$ is Fermi
energy\cite{FL-lifetime}. However, in a finite system all
single-- and many--particle states are discrete. In this case,
quasiparticles may be viewed as wave--packets constructed of
such states, the packet width being determined by the
quasiparticle lifetime in an infinite system:
$\delta\epsilon\simeq\gamma(\epsilon)$.
In this paper we attempt at making the relation between quasiparticles and
many-particle states more precise, and find that at different
energies it has different meaning.

Conventionally, quasiparticles are well defined provided
$\gamma(\epsilon)\ll\epsilon$.
However, to resolve quasiparticles in a mesoscopic system,
a more stringent condition is required: $\gamma(\epsilon)<\Delta$,
the quasiparticle level spacing.
As an example, we consider quasiparticle
peaks in tunneling conductance of a quantum
dot\cite{Sivan94a,Ralph96}. The peaks are observed in non--linear
conductance at certain bias, and
are interpreted in terms of energy dependence of the
quasiparticle tunneling density of states (DOS),
so that each peak corresponds to a ``quasiparticle
state,'' and its width measures the lifetime of the state.
Below, in our discussion of the width in terms of constituting
eigenstates, we ignore any contributions to the quasiparticle
decay due to finite escape rate, phonons, etc.\cite{Kamenev96},
and consider an isolated Fermi-liquid.

The meaning of quasiparticle lifetime needs clarification:
strictly speaking, since quantum dot is a finite system, any
many--particle eigenstate gives rise to an infinitely narrow
conductance peak. However, we will see that only a small
fraction of those states have significant overlap with
one-particle excitations, and thus can be detected by a finite
sensitivity measurement. Under certain conditions, these strong
peaks group into clusters of the width $\sim\gamma(\epsilon)$,
that can be interpreted as quasiparticle peaks.

Before discussing possible regimes let us review the GR
approach. Recently Sivan et al.\cite{Sivan94}, adopting the
quasi--particle picture to a finite size geometry and relying on
the earlier work\cite{Altshuler82} on electron--electron
scattering rate in diffusive conductors found that
  \be
\gamma(\epsilon)
\approx\Delta\cdot \left(\epsilon/g \Delta \right)^2\ ,\qquad
\epsilon< g\Delta\ ,
\label{sia}
   \ee
where $\Delta$ is the mean single--particle level spacing near
Fermi level and $g\gg 1$ is the dimensionless conductance, for a
finite system defined by $g=E_{c}/\Delta$, where $E_c$ is the
Thouless energy (inverse time of diffusion through the system).
The decay rate (\ref{sia}) is much larger than in a clean Fermi-liquid,
however, at $\epsilon<E_c$
one has $\gamma(\epsilon)\ll \Delta$, implying that the
quasiparticle states can be resolved.

However, the GR can be used to evaluate lifetime only
when the density of final states is sufficiently large, so that
the GR decay rate is larger than the level spacing of
final unperturbed states. Otherwise, the GR will not  give
the decay rate, but rather just a first--order perturbation
correction to the energy of a given eigenstate. It is important
to realize that in our problem, since a quasiparticle decays
into three quasiparticles, the density of relevant final states,
$\nu_3(\epsilon)=\epsilon^2/(2\Delta^3)$, is much smaller than
that of all many--body states. The interaction matrix element $V$ in the GR leading to
Eq.~\ref{sia} is of the order of $\Delta/g$ (see below), which
should be compared to the three-particle level spacing
$1/\nu_3$. Therefore, the GR is not applicable unless
$\epsilon > \epsilon^*=\Delta\sqrt{g}$. Note that, since
$\epsilon^*\gg\Delta$, there are many states whose lifetime is
not given by  GR.

At $\epsilon\ll \epsilon^*$, when matrix elements are much smaller
than the spacing $1/\nu_3$, the quasiparticle states do not decay:
they are just slightly
perturbed one--particle states. Hence they produce strong and very narrow
conductance peaks, some of which may have weak satellites
due to coupling to many--particle states. As $\epsilon$ approaches $\epsilon^*$
from below, the number of the satellites rapidly increases, and
at $\epsilon\gg\epsilon^*$, they altogether form finite width
peaks, well described by the GR.

For a quantitative description of the whole interval
$0<\epsilon<E_c$ (including the vicinity of $\epsilon^*$),
it is both interesting and instructive to
explore the analogy of this problem with the Anderson
localization.  This will be the goal of this paper.

Extension of the traditional localization problem to few
interacting particles has received much attention recently. The
study of the two-particle case, started by
Dorokhov\cite{Dorokhov}, was further advanced by
Shepelyansky\cite{Shepelyansky94}, Imry \cite{Imry95} and
Pichard et al.\cite{Weinmann96}, with extensions to more
particles. Some of the energy scales encountered below, such as
$V$ and $1/\nu_3$, have already been discussed in the context of
those papers. In contrast, here  we deal with the states extended
throughout the whole finite system. The localization we consider occurs
in the Fock space of many-body states, rather than in the real space.

\smallskip\noindent
{\it Distance in Fock space\/} is a concept we introduce to measure
closeness of many--body states.
Consider a generic two--body interaction Hamiltonian
in a secondary quantized form
   \be \label{H}
{\cal H}_0+{\cal H}_1 = \sum_{\alpha} \epsilon_\alpha c^+_\alpha c_\alpha
+\sum_{\alpha\beta\gamma\delta} V^{\alpha\beta}_{\gamma\delta}
c^+_\gamma c^+_\delta c_\beta c_\alpha. \nonumber
   \ee
The many--body problem is formulated in the Fock space, by
choosing as a basis $\{\Psi_N\}$ -- the Slater determinants
constructed out of the $N$--particle Fermi vacuum $|N\ra$:
  \begin{equation}
\Psi_N=c^+_{\alpha_{2m}}\ldots c^+_{\alpha_{m+1}}
c_{\alpha_{m}}\ldots
c_{\alpha_1}|N\ra \ .
  \end{equation}
Any state $\Psi_N$ can be represented as a string with entries 1
and 0 labeling the single particle states which do or do not
participate in $\Psi_N$, respectively.  Let us define the
distance between two states $\Psi_N$ and $\Psi'_{N}$ as the
number of positions in which the corresponding strings differ.

Thus defined, the distance in the Fock space can be used to map
the lifetime problem to an appropriate
Anderson localization problem.  It is
useful to think of the states $\Psi_N$, the eigenstates of
${\cal H}_0$, as of ``site orbitals'' in the Fock space, each having an
on--site energy
$\epsilon_{\alpha_{2m}}+...+\epsilon_{\alpha_{m+1}}+
|\epsilon_{\alpha_{m}}|+...+|\epsilon_{\alpha_1}|$.
These sites are interconnected by the interaction ${\cal H}_1$, which we
think of as  hopping in the Anderson problem (the diagonal
part of ${\cal H}_1$ is added to ${\cal H}_0$ by using the Hartree--Fock
method). The point is that the two--body interaction matrix element
$\la\Psi_N|{\cal H}_1|\Psi'_{N}\ra$ is non--zero only if the distance
between the states $\Psi_N$ and $\Psi'_{N}$ equals 0, 2 or 4.
We construct a network in the Fock space by connecting
all orbitals $\Psi_N$ which are at a distance 2 from each other.

Below we study the localization problem on this network.  We show that
the energies at which the GR is relevant correspond to
states extended over the network, whereas at lower energies the
hopping does not form enough resonances between neighboring
orbitals, resulting in localization of eigenstates near original
sites.  Unlike in other localization problems, as the energy is
lowered, localization develops very gradually, within a large
critical region, $\epsilon^{**}<\epsilon<\epsilon^*$, where
$\epsilon^{**}\simeq\epsilon^*/\sqrt{\ln g}$ and
$\Delta<\epsilon^{**}<\epsilon^*<E_c$. This is a result of the
particular structure of the network in the Fock space.  Above the
localization threshold (and beyond the critical region)
we recover the GR
picture\cite{Sivan94} with finite width quasiparticle conductance
peaks.  Each of them is formed by a huge number of eigenstate
peaks, and even a small external broadening will smear this fine structure.
In the localized regime
$\epsilon<\epsilon^{**}$ the single-particle density of states
will consist of isolated $\delta-$function peaks. In the critical
region, $\epsilon^{**}<\epsilon<\epsilon^*$, the peaks have
irregular structure which turns into simple Lorentzian above
$\epsilon^*$.

The meaning of localization in the Fock space is that a
localized state is practically identical to a single--particle
excitation (or a superposition of very few quasiparticle
states). The energy of each of the constituent quasiparticles
represents a good quantum number, whereas for the extended
delocalized states only total energy is conserved. The
transition is of the Anderson type because the two-body
Hamiltonian is {\it local} in the Fock space: it couples only
the orbitals of similar quasiparticle content.

It is worth remarking that the hierarchical
organization of many--particle states proved to be a useful
picture in studies of compound nuclear reaction
rates\cite{Blann75}. The ``doorway states'' introduced in the
nuclear reaction studies, although serve a different purpose,
are related to our network construction.

\smallskip\noindent
{\it Hopping over the network in the Fock space:\/}
The two--body interaction matrix elements in (\ref{H}) are given by
  \be
\label{V}
V^{\alpha\beta}_{\gamma\delta}=\int\!\!\!\!\int\!\!dx dx'
V(x-x')\psi^*_\delta(x)\psi^*_\gamma(x')  \psi_\beta(x)\psi_\alpha(x')\, .
  \ee
To evaluate the matrix elements, let us consider diffusive
disorder and a short range interaction, $V(x-x')=\lambda \Delta
{\cal V} \delta(x-x')$, where ${\cal V}$ is the volume, and
$\lambda\approx 1$ is the dimensionless interaction strength.
For $\alpha\ne\beta\ne\gamma\ne\delta$,
$V^{\alpha\beta}_{\gamma\delta}$ is a random quantity with zero
average. The root--mean--square $V$ may be evaluated\cite{foot2},
e.g., by using the diagram shown in Fig.~1.a.
In the absence of time reversal
symmetry we obtain
  \be \label{V-mean-square-root}
V=\lambda b_d {\Delta^2\over E_c}\, ; \quad
b_d^2=\frac{2}{\pi^2}
\sum\limits_{m\neq 0}\frac{\gamma_1^2}{\gamma_m^2} \, ,
  \ee
where $\gamma_m$ are eigenvalues of the diffusion operator. (By
definition, $E_c=\gamma_1$.) In deriving
(\ref{V-mean-square-root}) we assume small single--particle
energies: $\epsilon_{\alpha(\beta,\gamma,\delta)}\le E_c$. The
magnitude of $V$ decreases algebraically when the differences
between the single--particle energies exceed $E_c$; below we
shall ignore such contributions.

The network in the Fock space organizes all states in a
hierarchy. Let $|N-1\ra$ be the ground state of $N-1$ particles.
The states $\Upsilon^\alpha = c^{+}_{\alpha}|N-1\ra$
representing one particle added in the state $\alpha$ form the
first generation of the hierarchy (see Fig.~1.b).
The states $\Upsilon^{\alpha
\beta}_\gamma = c^{+}_{\alpha} c^{+}_{\beta} c_{\gamma}|N-1\ra$,
representing two particles and one hole, form the generation 3.
Similarly, the generation 5 is formed by
$\Upsilon^{\alpha\beta\gamma}_{\lambda\mu}= c^+_\alpha c^+_\beta
c^+_\gamma c_\lambda c_\mu |N-1\ra$, etc. The two--body
interaction ${\cal H}_1$ couples only the states of near
generations, so that any given state from  generation $2n-1$ is
connected only to states from  generations $2n+1$, $2n-1$,
or $2n-3$.
This implies that connected states are a distance 2 from each other.

Consider now a state of generation 1, with an on--site energy
$\epsilon$. The DOS in generation 3 accessible by ``hopping,''
having the same energy is
$\nu_3(\epsilon)=\epsilon^2/(2\Delta^3)$.  For higher-order
generations (more excited quasi-particles), the DOS rapidly
increases: for the $(2n+1)^{st}$ generation (where $n<
n_{max}\approx \sqrt{\epsilon/\Delta}$) it goes as
$(\epsilon/\Delta)^{2n}/(2n)!$. However, we should focus only on
those states of  generation $(2n+1)$ which are {\it directly
accessible} from a {\em given} state of  generation
$(2n-1)$. The density of such states is much smaller, and is
given by $\nu_{2n+1}=\nu_3/n$.  We note that from a state in the
generation $(2n-1)$ it is also possible to hop to some states of
the same generation, and to some states of the previous
generation $(2n-3)$. Respectively, DOS associated with these
processes is $\sqrt{\nu_3/\Delta}$ and $n(n-1)(2n-3)/\Delta$.
For $n\ll n_{max}$ the number of such hopping processes is
parametrically smaller than the number of states in the next
generation accessible by hopping. We thus obtain a picture which
is quite close to that of a Cayley tree: each ``site'' of the
$(2n-1)^{st}$ generation branches out to $K_n$ sites of the next
generation. (The number of couplings to the sites of the same
or of the previous generations is much smaller, and thus can be
ignored\cite{foot3}.) The branching number is given by
integrating the effective DOS over the energy interval $E_c$
where the hopping parameter $V$ is energy independent.  We obtain
the branching number
  \be \label{K}
K_n\approx  g^3/(6n)\, .
    \ee
Note that $K_n$ depends on the hierarchy level of the tree.

Decreasing the branching number $K_n$ with increasing $n$ makes
the network effectively finite. In order to simplify the
consideration, below we will consider an {\it infinite Cayley
tree} with {\it constant branching number} $K=K_{1}=g^{3} /6$.
More realistic model, taking into account finite size of the
tree and $n$-dependence of the branching number, will be
considered elsewhere \cite{tobepublished}.
Qualitatively, as a result of the
finiteness of the network, the localization transition we find
in the infinite network will be smeared to a crossover. Up to
the finite size smearing effects, all the results obtained for
the infinite Cayley tree will remain intact.

The model we are interested in was solved by Abou--Chacra,
Anderson and Thouless\cite{Chacra73}. They considered
localization on a Cayley tree with the on--site energies from a
uniform distribution in the interval $[-W,W]$, and constant
hopping amplitude $V$.  By studying fix--points of the mapping of
self-energies computed recursively using the hierarchy of the
Cayley tree, it was found that delocalization occurs at $V\simeq
W/(K\ln K)$, where $K$ is the tree branching number. This result is
in apparent disagreement with the Anderson criterion of
localization, $V\simeq W/K$, obtained by comparing the hopping
amplitude with the spacing of the on--site energies in the nearest
neighbor shell.

To understand the significance of $\ln K$, instead of following
the exact solution\cite{Chacra73}, let us study the statistics of
resonances of the tree sites appearing due to hopping.
Starting at a site of the first generation, the amplitude
(at energy $\epsilon$)
connecting this site to a given site at the
$(2n+1)^{st}$ generation, in
lowest order in $V$ is given by
  \be
\label{A}
A_n =\prod^n_{i=1} {V\over \epsilon-\epsilon_i}\, .
  \ee
Let us consider the probability, $p(n,C)$, that this amplitude
is significant, namely that $|A_n|$ exceeds given finite $C$. We
define $\ln |A_n|\equiv n\ln {V\over W}+Y_n$, where
$Y_n\equiv\Sigma^n_{i=1} y_i$ and $y_i\equiv\ln
\big|{W\over\epsilon-\epsilon_i}\big|$. Assuming that the
relevant on--site energies are uniformly distributed in the
interval $-W\le\epsilon-\epsilon_i\le W$, the probability
distribution of $y_i$ is $\tilde P(y_i)=\exp\{-y_i\}$ where
$0\le y_i<\infty$. Fourier transforming $\tilde P$, taking the
$n^{th}$ power and Fourier transforming back, we obtain the
distribution function of $Y_n$, $(0\le Y_n< \infty)$. Since
$|A_n|=({V\over W})^n \exp\{Y_n\}$, the distribution function of
$|A_n|$ will be
  \be
\label{adistr}
P(|A_n|)={(V/W)^n \over (n-1)!}\,  {1\over |A_n|^2}
\left[\ln \left( |A_n|\left(W\over V \right)^n\right)\right]^{n-1}\, .
  \ee
Thus the probability $p(n,C)=\int^\infty_C dA\, P(A)$ for $W\gg V,\,
C\approx 1$ is given by
  \be
\label{p}
p(n,C)\approx{1\over (n-1)!} {1\over C\,\ln\left( C({W\over V})^n \right)}
\bigg[{V\over W} \ln \left( C\bigg({W\over V}\bigg)^n \right) \bigg]^n.
  \ee
The probability that {\em none} of the $K^n$ trajectories connecting a
site in the first generation to sites in the generation $(2n+1)$ carries a
large amplitude is given by
  \be
\label{f}
(1-p(n,C))^{K^n}\equiv \exp (-f_n)\, ,
  \ee
where for $p(n,C)\ll 1$,  $f_n\approx K^n p(n,C)$.
From Eq.~\ref{p} for $n\gg 1$ one obtains
  \be
\label{f1}
f_n\approx {1\over\sqrt{2\pi n}\, C}\,{1\over \ln {W\over V}}
\bigg[{K V e\over W}\ln {W\over V} \bigg]^n  \, .
  \ee
If $f_n$ increases at large $n$, then, eventually, at
higher generations one has  $f_n\gg 1$, i.e., strong coupling
to generation 1. A transition to the  localized phase
takes place when the expression in the
square brackets in Eq.~\ref{f1} assumes
the value 1, which gives the criterion quoted above.

To apply this result to our problem, we replace
$V^{\alpha\beta}_{\gamma\delta}$ by $V$ and approximate the density of states
in the generation $(2n+1)$, accessible from a state (of energy
$\epsilon$) in the generation $(2n-1)$, to be uniform in the interval
$[\epsilon-W,\epsilon+W]$, $V<W<\epsilon$, and equal to
$K/(2W)=\nu_3(\epsilon)$. We then find that the transition
occurs at the energy\cite{foot4}
  \be
\label{star}
\epsilon^{**}\approx   (\lambda b_d 2 e)^{-1/2}
\cdot\sqrt{\Delta  E_c\over \ln g}\,\,  .
  \ee
At  energies just above $\epsilon^{**}$
the first generation is not very well connected with the next few
generations ($f_n< 1$ for small values of $n$). The condition for {\it all}
generations to be well connected with the generation 1 is
$f_1=K V/(W C)>1$.
This coincides with a naive implementation of the
Thouless criterion \cite{Thouless77}, yielding the second energy scale
  \be
\label{star2}
\epsilon^*=\left( \frac{\lambda b_d}{C}\right)^{-1/2}
\sqrt{\Delta E_c}\,\,  .
  \ee
A convenient expression for $f_n$ is, then,
\be
\label{f3}
f_n(\epsilon)\approx {1\over \sqrt{n}}
\bigg({\epsilon^{**}\over\epsilon^*}\bigg)^2
\bigg({\epsilon\over\epsilon^{**}}\bigg)^{2n}\, .
\ee

Let us discuss the meaning of the various regimes.  In the
localized phase ($\epsilon< \epsilon^{**}$) the first generation
is weakly connected with the rest of the network.  Therefore, at
such energies the exact many--body states are close to Slater
determinants, or
to  superpositions of few determinants.
A mathematical description of a single particle injection
into a dot involves projecting a single--particle state onto exact
eigenstates of the system. In the localized phase each single--particle state
will have a significant overlap with one (or few) exact eigenstates,
producing a few resolved $\delta$--function peaks in the spectrum of the
single--particle DOS.
At $\epsilon>\epsilon^*$ all generations are well connected.
Due to the huge density of multiparticle states, the states of
generation 1 can be thought of as being effectively
well coupled to the continuum.  This justifies
the GR result, Eq.~\ref{sia}, in this energy
range\cite{tobepublished,Imry96}.
Each single particle peak associated with  generation 1 is replaced by a
cluster of a large number of many-particle peaks, altogether forming a
Lorentzian envelope.
For $\epsilon<E_{c}$,
the width of the envelope is less than $\Delta$, and thus the
``quasiparticle states'' can be resolved in, e.g., transport
measurements \cite{Sivan94}.  For
intermediate energies $\epsilon^{**}<\epsilon<\epsilon^*$ there
are still many peaks in a cluster.  However the probability
that a particular generation is represented in a given cluster is
small.  As a result the widths of clusters as well as the
shapes of their envelopes will strongly fluctuate from peak to peak.

Following from our discussion, there are some interesting
implications to the relation of Quantum Chaos and Anderson
Localization. As long as quasiparticle peaks can be resolved,
the many--particle spectrum is not trully chaotic in the
Wigner--Dyson sense. (For example, from our analysis, different
quasiparticle states ``do not talk to each other.'') Another
observation is that an extended state in the many--body problem
is not necessarily ergodic, but can be extended over a
small fraction of the whole space (in our case, over a subtree
of the Cayley tree).

\bigskip
We acknowledge useful discussions with H.~Feshbach, Y.~Imry,
V.~E.~Kravtsov, S.~Levit, J.~L.~Pichard, U.~Sivan, A.~Stern and
H.~A.~Weidenmuller. The work at Weizmann Institute was supported
by the U.S.-Israel Binational Science Foundation, the Israel
Academy of Sciences and Humanities, the German-Israel Foundation
(GIF). The work of L.~L. was supported primarily by the MRSEC
Program of the National Science Foundation under award number
DMR 94-00334.



\begin{figure}[hbt]
\psfig{file=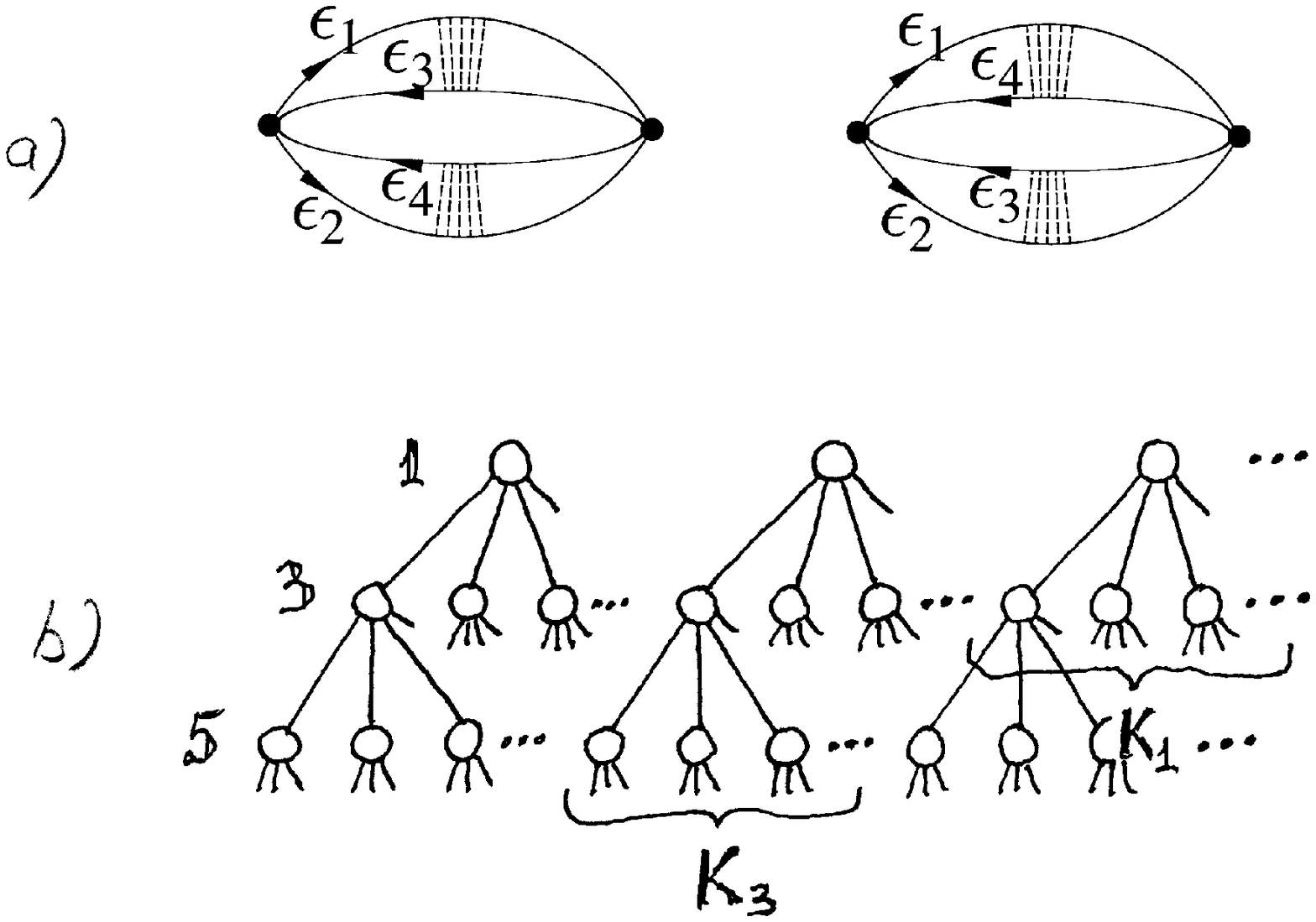}
\caption{
   {\it (a)}
The diagrams used to evaluate the mean square of
the matrix element in Eq.~\protect{\ref{V} }.
   \hskip1mm {\it (b)}
Schematic representation of the Cayley tree in the Fock space of
many--body states. Different generations are shown.
}
\label{fig1}
\end{figure}


\begin{thebibliography}{99}
\bibitem{foot2} We have learned of a similar calculation by
Ya.~M.~Blanter, cond--mat/9604101.
\bibitem{foot3} Within this
approximation a site of generation $(2n+1)$ is connected to a single site
of generation $(2n-1)$.
\bibitem{foot4}Note that  $\epsilon^{**}$ is independent of $C$.
We also note that the density of sites available from a given side in a
previous generation depends on $\epsilon$ (cf. expression for $\nu_3$). Hence
unlike the above treatment it is not rectangular. It easy to check that our
results, say for $\epsilon^{**}$, are fairly robust against the choice of the
value of $K$, $W$ (or, in general, shape of the DOS),
as long as we keep the ratio $K/(2W)=\nu_3(\epsilon)$ fixed.
\end{thebibliography}
\end{document}